\documentclass[%
 reprint,
 longbibliography,
superscriptaddress,
 amsmath,amssymb,
 aps,
]{revtex4-2}

\usepackage{graphicx}
\usepackage{dcolumn}
\usepackage{bm}
\usepackage{tabularx}
\usepackage[raggedrightboxes]{ragged2e}
\usepackage{subfig}
\usepackage{amsmath}
\usepackage{multirow}
\usepackage{xcolor}

\newcommand{\red}[1]{\textcolor{black}{#1}}

\preprint{APS/123-QED}

\begin{document}

\title{Comparing Large Language Models for supervised analysis of students' lab notes}

\author{Rebeckah K. Fussell}
\affiliation{Laboratory of Atomic and Solid State Physics, Cornell University, Ithaca, New York 14853, USA}
\author{Megan Flynn} 
\affiliation{Laboratory of Atomic and Solid State Physics, Cornell University, Ithaca, New York 14853, USA}%
\affiliation{Department of Computer Science, Cornell University, Ithaca, New York 14853, USA}
\author{Anil Damle}
\affiliation{Department of Computer Science, Cornell University, Ithaca, New York 14853, USA}
\author{Michael F.J. Fox}
\affiliation{Department of Physics, Imperial College London, London, UK}
\author{N. G. Holmes}
\affiliation{Laboratory of Atomic and Solid State Physics, Cornell University, Ithaca, New York 14853, USA}

\date{\today}
\begin{abstract}
\red{Recent advancements in large language models (LLMs) hold significant promise in improving physics education research that uses machine learning. In this study, we compare the application of various models to perform large-scale analysis of written text grounded in a physics education research classification problem: identifying skills in students' typed lab notes through sentence-level labeling. Specifically, we use training data to fine-tune two different LLMs, BERT and LLaMA, and compare the performance of these models to both a traditional bag of words approach and a few-shot LLM (without fine-tuning).} We evaluate the models based on their resource use, performance metrics, and research outcomes when identifying skills in lab notes. We find that higher-resource models often, but not necessarily, perform better than lower-resource models. We also find that all models estimate similar trends in research outcomes, although the absolute values of the estimated measurements are not always within uncertainties of each other. We use the results to discuss relevant considerations for education researchers seeking to select a model type to use as a classifier. 

\end{abstract}
\maketitle
\section{\label{sec:introduction}Introduction}
Open-source large language models (LLMs) have the potential to transform research methods in physics education research (PER). Older natural language processing (NLP) methods have been in use in PER for many years; common methods include bag of words~\cite[e.g.,][]{fussell_machine_2022, geiger_developing_2022, nakamura_automated_2016, wilson_classification_2022, odden_thematic_2020, odden_how_2021, mariegaard_identification_2022, bralin_analysis_2023, wulff_computer-based_2021, rosenberg_combining_2021, ullmann_automated_2019, young_exploring_nodate}, a method in which model inputs are an unordered collection of individual words, or statistical word vector embeddings~\cite{sherin_computational_2013}, a method in which words are assigned a vector such that synonyms have similar directions and magnitudes.
Cutting-edge LLMs are purported to offer significant improvements over these methods. Researchers have begun investigations into using proprietary models for PER, including IBM's Watson~\cite{campbell_using_2022, campbell_evaluating_2024} and ChatGPT~\cite[e.g.,][]{kieser_educational_2023}. 
Unfortunately, full details of these models have not been released publicly and secure locally-licensed versions may not be available to all researchers. Privacy and security are of particular concern to physics education researchers who must protect human subjects data.
Thus open-source LLMs that allow the student data and fine-tuned model to remain local on a private and secure computer, such as BERT~\cite{devlin_bert_2019} and the even larger LLaMA models~\cite{touvron_llama_2023}, are an advantageous and under-researched methodology for PER. 
Each machine learning algorithm offers distinct benefits and tradeoffs, however, and so it is crucial to evaluate the models being implemented and to quantify the uncertainty when making claims about students~\cite{fussell_method_2024}. 

This study is part of a larger project that aims to identify experimental skills in students' written lab notes, such as if students make comparisons between their data or propose methodological iterations. Our aim is ultimately to measure the frequency with which students employ these skills in their lab notes across lab units, institutional contexts, and teaching interventions, and to scale up analysis to achieve broader representation of diverse populations of physics students~\cite{kanim_demographics_2020}. Here we evaluate the extent to which large language models support that aim. 

In this study, we evaluate the effectiveness of different machine learning algorithms, including LLMs, in the context of an analysis that makes claims about students; namely, to measure the prevalence of certain experimental skills in students' lab notes across the semester (described below as the ``Research outcomes'' dimension of comparison). Our data source, student lab notes, has been used before in PER~\cite[e.g.][]{stein_confirming_2018, stanley_using_2017, holmes_developing_2020} and is a ubiquitous type of assignment in physics departments~\cite{hoehn_framework_2020,subcommittee_of_the_aapt_committee_on_laboratories_aapt_2014}. \red{In particular, our data source consists of lab notes that students type up during the lab period.} Thus lab notes serve as an interesting, common, word-based (instead of symbol/number-based) physics data source for future multi-institutional studies that apply LLMs. 

\subsection{LLMs in PER}


Current methods in \red{using NLP to apply labels to} student data can be split into two groups: \red{(i) methods} that apply pre-trained machine learning models to data without re-training the model for the specific task (though there may be some engineering of how the model is prompted) and (ii) methods that use task-specific training data to train a model to reproduce a coding scheme. For a deeper overview of publications in Physical Review Physics Education Research or the Physics Education Research Conference proceedings that use machine learning methods, we direct the reader to~\cite{bralin_mapping_2024}.

LLMs are used often in the first category of studies, where \red{pre-trained} LLMs perform a task such as grading student work~\cite[e.g.,][]{kortemeyer_toward_2023, kortemeyer_grading_2024, kortemeyer_assessing_2024} or completing physics coursework or concept inventories~\cite[e.g.,][]{kortemeyer_could_2023, polverini_performance_2024} using only researcher-created prompts. \red{This type of methodology is referred to as ``zero-shot'' or ``few-shot'', where either zero examples or only a few task-specific examples are provided in the prompt, and the model itself is not altered.} This approach relies on careful attention to emerging prompt-engineering techniques~\cite[e.g.,][]{polverini_how_2023}. Other \red{studies that apply a pre-trained LLM without re-training} include exploration of using ChatGPT to simulate concept inventory data~\cite{kieser_educational_2023}, a study of students evaluating the scientific accuracy and linguistic ability of ChatGPT~\cite{dahlkemper_how_2023}, and an analysis of how ChatGPT reflects student misconceptions about physics~\cite{wheeler_chatgpt_2023}. 

The second category of studies involve using a human-coded dataset to train a model to code an education dataset. This method has a longer history in PER~\cite[e.g.,][]{wilson_classification_2022, nakamura_automated_2016, geiger_developing_2022, munsell_using_2021, ruf_comparison_2024} and connects more with familiar PER methods because it involves coding datasets with a theoretically-motivated coding scheme. \red{In machine learning, ``supervised'' refers to methods that use labeled data to train a model to predict the labels for a new data set.} The training for these supervised methods can either be from scratch (using a randomly initialized, untrained model) or by using a pre-trained model allowing the parameters of the model to adjust as the model is exposed to the training data (this process is commonly referred to as ``fine-tuning'')~\cite[e.g.,][]{campbell_evaluating_2024}. LLMs have rarely been used for this purpose, but are a promising option to improve performance on supervised classification tasks over ``traditional'' machine learning algorithms, which are more common in the past literature. 

One recent study directly compared the performance of a bag of words model (trained with about 280 data points) to a \red{few-shot} ChatGPT model (with three or fewer examples) and found that the bag of words model performed better for their binary classification task of assessing students' concept use in a physics problem~\cite{kieser_david_2024}. In this study, we expand on Kieser et al.'s comparison by using training data to fine-tune multiple LLM classifiers, and comparing the performance of these trained LLM classifiers to both a ``traditional'' bag of words approach and a \red{few-shot} LLM approach. 


The Kieser et al. study suggests that it will not always be the case that the largest, most recently released LLM is the best choice PER questions. One reason for this is the increased skill and resources involved in setting up a more recent, larger LLM from open-source code. Another reason is that LLMs are all different from each other, in ways that are both known and unknown. Different LLMs have different architectures and are trained with different methods and with different data sets. As a result, it is challenging to make general statements that compare the performance of different models. Tasks that are particularly of interest to PER---such as identifying skills in lab notes---are not necessarily similar to the benchmark tasks reported in the papers introducing LLMs~\cite{devlin_bert_2019, touvron_llama_2023}. The aim of this paper, therefore, is twofold: first, to motivate PER researchers (including ourselves) to use more LLMs in studies that use supervised analyses to automate coding of education datasets, and second, to share insights into which LLMs may be more advantageous to choose for these analyses as well as insights into how to evaluate if a particular LLM is the right choice for a given study.  

\subsection{Dimensions of comparison}
\red{There are a great number of relevant dimensions to choosing a machine learning model for a PER study. In this paper, we use only open-source models that we can run locally in order to explore the options available to researchers that need to keep human subjects' data secure and private. In deciding how to compare a diverse set of model types for the purposes of this study, we narrow in on three dimensions: performance, resources, and outcomes.} 

Performance and resources can be thought of as a trade-off. Performance may improve as LLMs are updated and as they increase in size (measured by number of parameters). Yet there are costs to using larger models. These costs include time to train and hardware requirements. Working with cutting-edge LLMs also requires specialized programming knowledge\red{. Training a model requires choosing hyperparameters, which are variables (e.g. learning rate, number of iterations) that prescribe the model and training process. Knowledge of good hyperparameters may already exist for older models whereas selecting effective hyperparameters for newer models requires more tinkering.} 

The rationale for examining outcomes in types of models is more subtle. There is increasing evidence that the specific model used impacts the research outcomes (i.e., the quantitative estimates being made). Various forms of evidence are converging in support of a principle that we call ``cautious predictor bias'', a form of systematic uncertainty when using machine learning models for measurement (e.g. the frequency of a code in a student population). This uncertainty will tend to overestimate measurements at low measurement values and underestimate measurements at high measurement values~\cite{fussell_method_2024, kortemeyer_grading_2024,el-adawy_exploring_nodate}. Models are \red{trained to disincentivize making} extreme predictions that differ more with the ``truth value'' if the prediction turns out to be wrong. Additionally, researchers have seen evidence for an ``Anna Karenina principle'' that biases models to more often correctly analyze the work of high-performing students than low-performing students, because ``happy'' correct answers tend to be alike and ``unhappy'' incorrect answers tend to be unique in their own ways~\cite{schleifer_anna_2024}, though this principle may not apply to all models. 


\subsection{Research goals}

This study explores and compares various methods for automating text data classification in PER: bag of words, BERT, LLaMA (fine-tuned without a prompt), LLaMA (fine-tuned with a prompt), and \red{few-shot} LLaMA. For BERT, we \red{use} the uncased version of BERT Base~\cite{google_bert_bert-base-uncased_nodate}. For LLaMA, we \red{use} LLaMA 3 8B~\cite{meta_llama_meta-llama-3-8b_nodate}. \red{These models are discussed in more detail in Section~\ref{ml_methods}}. We \red{investigate} these methods across three dimensions: performance, resource use, and research outcomes. We contextualize the analysis in the following PER question relevant to our ongoing work identifying skills in lab notes: ``How does the prevalence of experimental skills change over the course of lab units in our introductory experimental physics course?'' 

\section{\label{sec:methods}Methods}
\subsection{Data sources}

Our data are lab notes from two semesters of students taking the introductory experimental physics lab course at Cornell University. \red{The students typed the lab notes themselves during class. The students may have included images in their lab notes, but all images were removed during our data pre-processing.} The data were collected in 2019 and 2022 (notably years prior to the broad emergence of ChatGPT, meaning student text was not generated through an LLM). The combined data set of lab notes consisted of 873 notes in total, with 58,369 sentences in total. 
The lab notes are associated with three distinct lab units: a pendulum lab (L1), an objects in flight lab (L2), and a Hooke's law lab (L3). L1 and L3 took place over two distinct two-hour lab sessions (each denoted a and b). L2 took place in a single two-hour lab session in 2022, and over two distinct two-hour lab sessions in 2019. Students worked in groups of 3 or 4 and submitted lab notes as a group.  Students were instructed to think of their lab notes as ``a stream of consciousness... to document what you were doing and why you were doing it at several time points throughout the lab.'' 
We took a sample of 205 lab notes from the larger combined dataset and coded this sample by hand. Of the 205 lab notes, 120 were from 2019 (20 for each of the six sessions) and 85 were from 2022 (17 for each of the five sessions). 

We processed all lab notes documents by converting them to a spreadsheet with a new sentence on each row. We used the python-docx package to automatically identify separate sentences in the document~\cite{noauthor_python-docx_nodate}. We used a coding scheme that identifies two experimental skills in lab notes (adapted from previous work~\cite[e.g.,][]{holmes_teaching_2015, stanley_using_2017, subcommittee_of_the_aapt_committee_on_laboratories_aapt_2014}): Comparisons of Quantities (QC) and Proposed Iterations (PI). 

The QC code is defined as: ``Students should apply data analysis tools quantitatively to make some sort of comparison (between data, best fit lines, predictions, etc.).'' The inclusion rules are that a sentence must have a) evidence of a comparison between two quantities $x$ and $y$ and b) evidence that the quantities $x$ and $y$ are used for data analysis (e.g. mean, standard deviation, citing a specific measurement). Comparisons between data and a prediction, such as a model or best fit line, are included in the QC code. Instruction about making quantitative comparisons was introduced in L1b. During L1b, students were explicitly asked to apply this skill when analyzing their lab data.


The PI code is defined as: ``Students should be able to suggest additional rounds of experimentation and choose appropriate improvements. Could be based on experimental evidence.'' The inclusion rules are that a sentence must a) propose an experimental choice in either future or present tense and b) have at least one word or phrase that is synonymous with either ``more measurements'', ``changes/improvements'', or ``future plan''. All lab activities included some form of prompt for iterating on their experimental designs, though this was most explicit in L1a and L1b.

The coding scheme is applied at the sentence level; that is, each sentence is scanned for evidence of the code and each sentence is labeled as either containing or not containing the code.
We evaluated inter-rater reliability of our human coding scheme in two rounds: a first round with 12 lab notes (575 sentences total) and a second round with 5 lab notes (564 sentences total). Combining these two rounds, the Cohen's kappa for the QC code was 0.82 and the Cohen's kappa for the PI code was 0.78. This corresponds to inter-human balanced accuracy scores (defined in more detail below) of 0.89 and 0.86, respectively. Because the human coder had access to the full context of the sentences, we assume these scores represent the theoretical maximum performance of a classifier model for these two codes. 

\subsection{Machine learning methods} \label{ml_methods}
We implemented three different types of models to analyze the prevalence of the two coded skills, QC and PI, in students' lab notes. Below, we discuss general setup details, then describe the specific details for each model.

\subsubsection{General setup}
We used the spreadsheets that had been split into sentences by python-docx along with the human codes for these sentences as training and validation data. Prior to loading data into models for training or validation, we removed student names and identifying information. In all models, each sentence was treated independently. Additional pre-processing was model-specific and is defined below.

Our data were imbalanced, with many more sentences that did not have the code than sentences that did have the code, so it was important to balance the training data (equal number of sentences with and without the code)~\cite{young_predictive_2021, fussell_method_2024}. For bag of words, we achieved this by specifying balanced class weight using the scikit-learn package. For BERT and LLaMA, we used a weighted random sampler to balance the training data. Thus, for each model, batches of training data were sampled such that, on average, half the sentences in the batch would contain the code and the other half would not contain the code. 


\subsubsection{Bag of Words}
To prepare the sentences for the bag of words model, we applied the following text pre-processing protocol (similar to that used in our previous work~\cite{fussell_method_2024}): i) split all words in each response into individual words (often called tokens), ii) fix contractions (for example, ``you're'' becomes ``you are''), iii) use the Word Net Lemmatizer from the Natural Language Toolkit python package~\cite{bird_natural_nodate} to combine words from the same family such as plurals and verb conjugates, iv) replace numbers with the tags `DEC' and `INT' for decimals and integers, respectively, v) encode certain punctuation symbols with relevant semantic meaning as unique tokens (e.g., the question mark, the colon, and the math symbols for addition, subtraction, multiplication, and division), and vi) remove all other forms of punctuation, such as periods. 

We then encode the modified sentences into a matrix in which each row represents a sentence and each column represents a unique token (words or punctuation symbol). As in our previous work~\cite{fussell_method_2024}, each entry in the matrix is a 0 or 1 indicating if that token is present or absent in the sentence. We used the logistic regression algorithms from the scikit-learn package in python~\cite{pedregosa_scikit-learn_nodate} with the following parameters: $\ell_2$ regularization, maximum iteration limit of 10,000, and balanced class weight. 

\subsubsection{BERT}
To prepare the sentences for the BERT model~\cite{devlin_bert_2019}, we applied the following pre-processing protocol: i) remove any trailing white space from the end of sentence, ii) tokenize the sentence such that each word and punctuation mark is in an ordered list with the `[CLS]' token added to the start and the `[SEP]' token added to the end of each sentence, iii) compute the maximum list (i.e., tokenized sentence) length in the training set and pad each list with zeros such that each list is of the same length, iv) store information about which portions of each list contain meaning and which are padded in a data structure known as an attention mask. 

From an NLP perspective, this pre-processing protocol is simpler than the bag of words protocol because there are no steps taken to fix linguistic ambiguities arising from contractions or spelling errors, numbers are not converted into generalized tokens that stand in for all integers or decimals, and no punctuation is removed. BERT was originally trained on sentences with similar levels of minimal pre-processing, so we kept a similar protocol to the original training. 

The BERT model is a general language model which was pre-trained on two tasks: predicting masked words in a sentence and next-sentence prediction. Masked word prediction is a supervised prediction task in which a word in a sentence is masked and the algorithm must use the surrounding sentence context to predict the hidden word. Next sentence prediction is a supervised prediction task in which the algorithm must use a sentence to predict the next sentence. 
To accommodate these text outputs, the size of the output layer of BERT is 768 features. To construct a classifier, we had to modify the output to be just the models' estimated probability that the input sentence contains the code. To build the classifier, we appended a 50-neuron hidden layer to the 768 feature output of BERT, and an output layer with two neurons. The two-neuron output layer covers the probability associated with the two possible outputs---that the sentence contains the code or it doesn't. 

The parameters in the BERT language model are pre-trained by the models' original developers~\cite{devlin_bert_2019}, while the parameters in our linear classifier layers were not pre-trained and were initially assigned random numbers. In a process known as fine-tuning, we used our training data as inputs and the human coders' labels as outputs. \red{This process adjusts all the classifier parameters, both the pre-trained BERT parameters and the added layers, to be suitable for identifying if the specific skill is present.} 

The BERT classifier computations were performed on an NVIDIA A6000 GPU. The batch size was 16 sentences at a time, and we trained for 3 epochs. We used the pytorch package~\cite{paszke_pytorch_2019}. 


\subsubsection{LLaMA}
To prepare the sentences for the LLaMA algorithm~\cite{touvron_llama_2023}, we followed a similar protocol to the BERT pre-processing. Similar to the BERT classifier, we balanced the training data by assigning a weight based on the human label such that there is equal probability that sentences with and without the code (as determined by the human coder) would be sampled in a batch. The batch size was 16, and we ran 5 epochs. 

As with BERT, our fine-tuning procedure adapted the LLaMA model for this classification task, as LLaMA was originally trained to predict the next token in a sequence. \red{We used the AutoModelForSequenceClassification object in the transformers package to specify that the output should be a \emph{classification} (that is, a label, rather than text as in a chat bot)~\cite{wolf_huggingfaces_2020}}. As with the BERT classifier, the model computations were performed on an NVIDIA A6000 GPU. This GPU has 48 GB of memory and LLaMA is a very large model. Simply loading LLaMA's \red{8} billion parameters with float16 precision uses 14 GB of memory, so a full fine-tuning to retrain all model parameters would be infeasible. We addressed this by implementing LoRA, a method that reduces the number of trainable parameters as we fine-tune the classification algorithm~\cite{hu_lora_2021}. 

We implemented the LLaMA classifier model both with and without a prompt. For the QC code, we used the following prompt: ``Analyze the sentence given. Does the sentence demonstrate a quantitative comparison between data, best fit lines etc?'' For the PI code, we used the following prompt: ``Analyze the sentence given. Does the sentence demonstrate: Iteration (Proposed) - Students should be able to suggest additional rounds of experimentation and choose appropriate improvements. Could be based on experimental evidence. Answer with just yes or no.'' 

In preliminary testing with a less-optimized version of LLaMA, we tested three levels of prompt for each of the two codes. All prompts began with ``Analyze the sentence given. Does the sentence demonstrate...''. In the first level, the prompt was completed with a brief, simple summary of the definition and most common examples. In the second level, the prompt was completed with the full definition from the coding scheme. In the third level, the prompt was completed with the full coding scheme entry, including the definition, inclusion rules, and examples. \red{The second and third level prompts also included an explicit instruction to answer with Yes or No.} We found in this initial testing that the first level prompt had the best performance for the QC code and the second level prompt had the best performance for the PI code. 

Additionally, we implemented a \red{few-shot} LLaMA classifier. \red{``Few-shot''} means that we performed no fine-tuning to the particular task of identifying skills in lab notes, such as by providing the model with human coded training data. \red{Instead, we prompted an unmodified LLama-3-8B model with each sentence individually. All prompts began with the same language: an abridged version of the coding scheme, instructions to answer with a Yes or No, followed by four examples. All prompts ended with the sentence for the model to code, in the same format as the examples. The LLama-3-8B model was unmodified thus the output of the model was text.} This approach essentially uses the LLaMA model as a general-knowledge chatbot, rather than a model trained with data to perform a particular task. If the first three words of the model's reply contained the word Yes or No, we converted the reply to an appropriate numerical format (Yes = 1 and No = 0) so that we could compare directly with the human codes. If neither word occurred in the first three words, we planned to treat this as the model saying that the sentence did not contain the code, however, this did not occur in our data collection.

\subsection{Analysis methods}
We provide a jupyter notebook to accompany the analysis in this article~\cite{flynn_rkfussellcomparing_llms_public_2024}.

\begin{figure}[t]
{\includegraphics[width=1\linewidth]{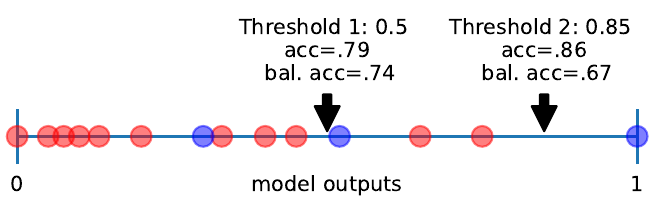}}
  \caption{A toy example demonstrating how the model output and threshold are used in conjunction to determine the model's classification of whether or not each sentence contains or does not contain the code. Red circles represent sentences that humans labeled as not containing the code and blue circles represent sentences that humans labeled as containing the code. If the threshold line is moved up or down, metrics like accuracy and balanced accuracy may change as sentences fall on different sides of the threshold line.~\label{toy_example}}\vspace{-1em}
  
\end{figure}
\subsubsection{Data sets}
Out of the 205 human-coded lab notes files, 22 were set aside as test data. These 22 lab notes files represent two lab notes from each lab session for each of the two semesters (six lab sessions in 2019 and five lab sessions in 2022). We refer to the test data as ``hold out'' data because we did not use these files while optimizing and selecting model hyperparameters. \red{If validation data are used to both tune a model's hyperparameters and report results, there is a risk of overestimating model performance~\cite{vabalas_machine_2019}. We created this hold out set to remove this risk}. 

We randomly split the remaining 184 lab notes files into training data (134 lab notes) and validation data (50 lab notes) in each model trial. The data were fed into the model as unordered individual sentences, but we ensured that sentences within the same lab notes file would not be split across the training and validation data sets. The model did not have access to the context of the individual sentences (e.g. the surrounding sentences). 

The validation data were used to tune the hyperparameters. \red{That is, before we settled on a set of hyperparameters we trained models with a handful of different combinations of hyperparameters and used the performance of these models on the validation data to inform our final selection of hyperparameters (also known as graduate student descent~\cite{grad_student_descent}). None of these validation data are used to evaluate model performance in our results.} 

\subsubsection{Resources Analysis Methods}
The research team had six NVIDIA RTX A6000 (48 GB) GPUs available to them through university computing resources, as well as Apple silicon processors on two of the researchers' personal computers. We chose our methods to align with these available resources. As we developed the code for each model type, we kept notes of the resources necessary to implement each model. These resources include level and type of human involvement in code implementation and development, necessary packages, and necessary hardware. 

We used the time package in python to measure how long it took to train each model with a training set of 5000 sentences. The timing does not vary significantly across repeated trials because each model runs a set number of computations for a given context length (length of each sentence) and number of sentences in the training set, and this does not vary across repeated trials within the same model. Adding a prompt to LLaMA increases the context length by a set amount and would thus increase the amount of time it takes to train/fine-tune.

\subsubsection{Performance Analysis Methods}

We compare the performance of models using three performance metrics: area under the curve (AUC), accuracy, balanced accuracy. 

To illustrate these metrics, we provide a toy example in Fig.~\ref{toy_example}. Each circle represents a sentence whose color indicates the human coding (red if humans labeled as not containing the code and blue if containing the code). The aim of a model is to indicate whether each sentence contains the code or not. The model computes an output (a number between 0 and 1) in the last stage of the calculation. Lower output values correspond to lower likelihood that the sentence contains the code and higher output values correspond to higher likelihood that the sentence contains the code (according to the model's calculations). If the model is any good (as well as the human coding, for that matter), the human coding should generally align with the model output (in terms of whether the sentence contains the code), especially in cases where the model is most confident: outputs close to zero and close to one. 

In computing these outputs, the model will tend to optimize for a particular ``threshold'', illustrated by the black arrows in Fig.~\ref{toy_example}. Outputs above the threshold are categorized as containing the code and outputs below the threshold are categorized as not containing the code. The threshold determines to what extent the model achieves a balance between minimization of false negative errors and minimization of false positive errors. For example, threshold 1 (at output value 0.5) in Fig.~\ref{toy_example} produces nine sentences correctly labeled as not having the sentence (true negatives), one sentence incorrectly labeled as not having the sentence (false negative), two sentences correctly labeled as having the sentence (true positives), and two sentences incorrectly labeled as having the sentence (false positives). Threshold 2 (at output value 0.85) produces 11 true negatives, two false negatives, one true positives, and zero false positives. The first threshold, therefore, has more true positives, while the second has fewer false positives. 

AUC is a threshold-independent measurement of how well the model separates data by category when computing the model output. In a balanced dataset, the AUC metric would be 0.5 if the model outputs are random. The AUC metric would be 1 if all sentences without the code have lower output values than all sentences with the code (in Fig.~\ref{toy_example}, this would occur if the dots were moved such that all red dots were to the left of all blue dots). 
In practice, perfect separation by condition is difficult to achieve. In the toy example in Fig.~\ref{toy_example}, the AUC metric would fall somewhere between 0.5 and 1. The six sentences with the lowest model output do not contain the code, a sign that the AUC metric will be better than random, while the next eight sentences are mixed, rather than separated. The AUC metric is useful, especially in cases of unbalanced datasets, because it gives threshold-independent information about the quality of a model's outputs. The AUC's threshold-independence, however, is also a limitation because, in reality, measurements will be made with respect to a model's threshold. Metrics that depend on the threshold, such as accuracy and balanced accuracy, are necessary to gain a full picture of the quality of result measurements.

Accuracy and balanced accuracy are metrics that use quantities from a confusion matrix. A confusion matrix (as in Table~\ref{confusion}) tells us the rate of true positives (TP), false positives (FP), false negatives (FN), and true negatives (TN) corresponding to a given threshold.  $N$ is the total number of sentences, such that $N = TP + FP + FN + TN$. We use the terminology of ``true'' and ``false'' according to standard terminology, though these should be interpreted as ``agrees'' and ``disagrees'' with human coding, respectively. 
Indeed, we have found sentences where post-hoc review of the sentences indicate that the model's prediction is more likely correct and the human coder either missed the code or applied the code incorrectly.

\begin{table}
\caption{The confusion matrix}\label{confusion}
\begin{tabular}{l|l|c|c|c}
\multicolumn{2}{c}{}&\multicolumn{2}{c}{Human coder}&\\
\cline{3-4}
\multicolumn{2}{c|}{}&Positive&Negative&\multicolumn{1}{c}{Total}\\
\cline{2-4}
\multirow{2}{*}{Model}& Positive & $TP$ & $FP$ & $TP+FP$\\
\cline{2-4}
& Negative & $FN$ & $TN$ & $FN+TN$\\
\cline{2-4}
\multicolumn{1}{c}{} & \multicolumn{1}{c}{Total} & \multicolumn{1}{c}{$TP+FN$} & \multicolumn{    1}{c}{$FP+TN$} & \multicolumn{1}{c}{$Total (N)$}\\
\end{tabular}
\end{table}

Accuracy is defined as $(TP+TN)/N$ and measures the fraction of sentences where both the model's prediction and human coder agree. 
This straightforward metric has advantages, particularly for interpretation, but is limited in cases where data are imbalanced. For example, if only 5\% of sentences contain a code, 95\% accuracy may be reported even if the model predicted that none of the sentences in the data set had the code. Accuracy also is threshold-dependent; in our toy model in Fig.~\ref{toy_example}, the accuracy is higher for threshold 2 than for threshold 1 (0.86 compared with 0.79 for the left-most).

Balanced accuracy may be more useful in cases of imbalanced data. Balanced accuracy takes the average of two other metrics: sensitivity and specificity. Sensitivity, also known as recall, is defined as $TP/(TP+FN)$ and measures the fraction of positive instances (according to the human coder) where both the model's prediction and human coder agree. In our toy example in Fig.~\ref{toy_example}, the sensitivity is higher for threshold 1, as fewer sentences with the code (as determined by human coders) would be falsely labeled as not having the code (blue circles below the threshold). Specificity is defined as $TN/(TN+FP)$ and measures the fraction of negative instances (according to the human coder) where both the model's prediction and human coder agree. In our toy example, specificity is higher for threshold 2, as fewer sentences without the code (as determined by human coders) are falsely labeled as having the code (red circles above the threshold). Balanced accuracy, therefore, seeks to balance these rates of false negatives (through specificity) and false positives (through sensitivity).

In our lab notes data there are far more sentences without the code than there are sentences with the code for both codes. This means that balanced accuracy is likely a better metric of the quality of the model than accuracy. The imbalance also means that balanced accuracy is likely to change more if the threshold value decreases than it does if the threshold value increases. The denominator on the sensitivity term (TP+FN) is smaller (few positive instances), so a slight threshold decrease can change the sensitivity term by a large amount while changing the specificity term by only a small amount. 




To calculate these performance metrics, we conducted five trials for each code and each model type. Each trial trained a new model (of the same type) with a different sample of training data and different random initial conditions. In all other respects the models were the same across trials. To compute the performance metrics, the five trained models were each applied to the same hold out test data and the metrics were averaged across trials. 

We also report performance metrics from the human inter-rater reliability data to contextualize the comparisons between models. We used a dataset comprised of 17 lab notes files coded by two separate individuals. We computed the inter-rater performance metrics by assigning one human coder to the role of ``model'' and the other to the role of ``human'' for the purposes of computing the performance metric. To represent the uncertainty of these inter-rater performance metrics we took eight lab notes files from the 17 at random and computed the performance metric, then repeated this 1000 times to calculate a mean and standard deviation. 




\subsubsection{Outcomes Analysis Methods}
For the outcomes analysis, we apply three trained models (bag of words, BERT, and unprompted LLaMA) to a new, large dataset consisting of 468 lab notes. Approximately 100 of these lab notes were hand-coded data and used in the training, validation, and test data described in the previous section. 
The remaining 368 lab notes were never hand-coded. 

The purpose of this analysis is to study how the outcomes of making education research measurements can vary across different model types. We apply the three trained models to this new, largely un-coded dataset to address the question ``How does the prevalence of experimental skills change over the course of lab units in our introductory experimental physics course?''. We report the results of this analysis as barcharts showing each model's estimate of the prevalence for each lab unit, along with our estimate of statistical and systematic uncertainty for each prevalence estimate. In doing so, we examine how models' estimates of prevalence are themselves dependent on the choice of model. \red{Full details of how we calculated statistical and systematic uncertainty are provided in Section~\ref{appendix}.}

\section{Results}
\begin{table*}[!htbp]
  \caption{Comparisons between the associated resources involved in each type of model evaluated in this study.~\label{tab:resources}}
    \begin{tabularx}{\textwidth}{ p{2.5cm} X  X  X  X  X}\hline\hline
      & \textbf{Bag of words Logistic Regression} & \textbf{BERT} & \textbf{LLaMA} & \textbf{Prompted LLaMA} & \textbf{\red{Few-shot} LLaMA}\\
      \hline
      \textbf{Parameters} & $\sim$4,000 (vocabulary size) & 110 million & 8 billion & 8 billion & 8 billion\\[0.5cm]
      \textbf{Time to train / fine-tune} & 2 s & 7 mins 16 s (A6000 GPU) & 6 hrs 21 mins (A6000 GPU) & 6 hrs 27 mins (A6000 GPU) & 0 - no training / fine-tuning\\[0.5cm]
      \textbf{Human involvement} & More human involvement in pre-processing & Can adapt code from examples online, pre-processing steps hurt performance & Hard to find example code online, examples may contain errors & Same as LLaMA, some creativity needed for prompt engineering & Some creativity needed for prompt engineering\\[0.5cm]
      \textbf{Packages} & scikit-learn & torch, transformers & torch, transformers & torch, transformers & torch, transformers\\[0.5cm]
      \textbf{Hardware} & CPU, any computer & GPU or Apple silicon chip (recommend at least 16 GB Memory)& minimum requirements met by a100 80GB GPU or A6000 GPU with LoRA & same as LLaMA & same as LLaMA\\[0.5cm]
      \red{\textbf{Number of labeled sentences}} & \red{8965 (average)} & \red{8965 (average)} & \red{8965 (average)} & \red{8965 (average)} & \red{4}\\[0.5cm]\hline\hline
    \end{tabularx}
\end{table*}
\subsection{Resources}
The resources required to implement bag of words, BERT and LLaMA are different (see Table~\ref{tab:resources}).
The size, as measured by number of parameters, of each model is of a different order of magnitude. Our bag of words model used logistic regression, which has one parameter for each vocabulary word in the training set (after our data preprocessing) plus one parameter for intercept scaling. For our dataset, this is typically around 4,000 parameters (varies slightly depending on the data sampled for the training set).
In contrast, BERT is an LLM with 110 million parameters. LLaMA is an LLM with 8 billion parameters, making it 80 times larger than BERT. 


The different sizes of these models correspond with vastly different time and energy costs associated with training/fine-tuning our classifiers. The bag of words model took 2 s to train on a standard CPU. The BERT model took 7 minutes and 16 seconds to fine-tune using an A6000 GPU; BERT would take longer to train on a chip with lower memory, such as a smaller GPU or an Apple silicon chip. The LLaMA algorithm took 6 hours and 21 minutes to fine-tune on an A6000 GPU. Similarly, the prompted LLaMA algorithm took 6 hours and 27 minutes. 

Working with each of these models involves different amounts of human involvement, which further informs the accessibility of each of the methods. LLaMA, for example, requires more expertise in setting up, debugging, and evaluating the code, which involves specialized packages ($torch$, $peft$, and $transformers$)~\cite{paszke_pytorch_2019, xu_parameter-efficient_2023, wolf_huggingfaces_2020}. Bag of words, in contrast, requires specialized NLP knowledge of how to pre-process the data, though the code itself is fairly straightforward to implement, particularly if leveraging online tutorials with $scikit-learn$. Prompted and \red{few-shot} LLaMA further require some prompt engineering. \red{Significantly more labeled sentences are needed to implement the four fine-tuned models (bag of words, BERT, LLaMA, and prompted LLaMA) whereas we only used four labeled examples in the Few-shot LLaMA model. We report the average number of labeled sentences used to train the fine-tuned models by multiplying the number of lab notes files in each training set (134) by the average number of sentences in a lab notes file (66.9).} 

Together, this resource use analysis indicates that bag of words is the most accessible method to implement without specialized computing resources, requiring the lowest amount of computing power. The LLaMA and Prompted LLaMA models are the most resource-intensive. BERT is relatively easy to implement, does not require as much expertise in specialized coding practices compared to LLaMA, and does not require expertise in pre-processing NLP practices necessary for bag of words. It is more resource-intensive than bag of words, but a computing setup with a GPU or Apple silicon chip and at least 16GB Memory is sufficient to meet the hardware requirements. Some higher-end commercially available personal computers already meet these hardware requirements.

\begin{figure*}[t]
  \subfloat[AUC]{\includegraphics[width=0.45\linewidth]{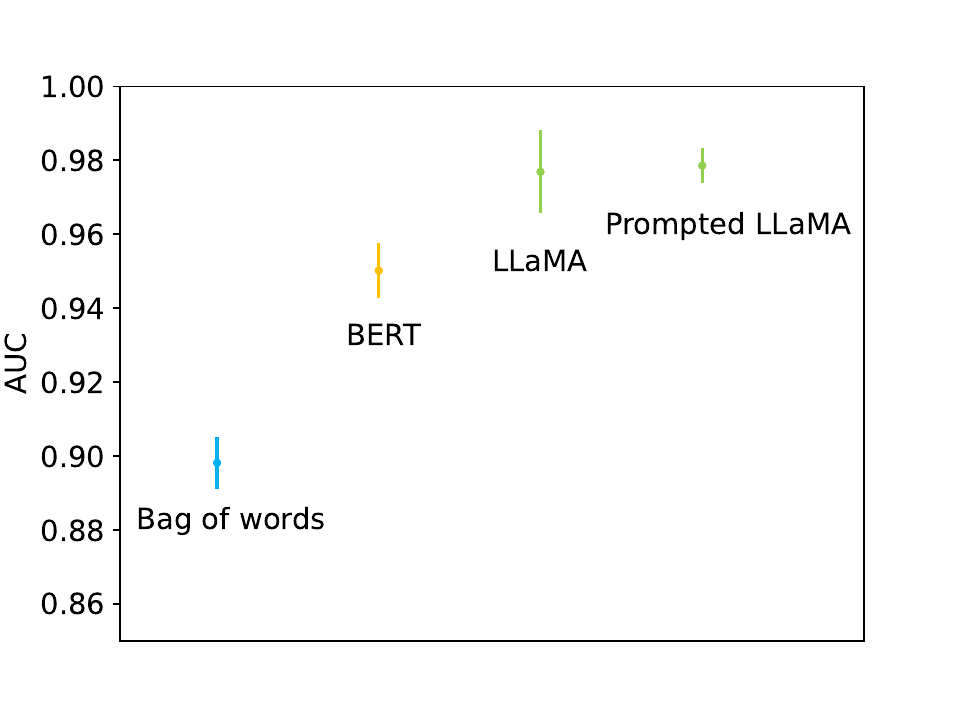}\label{QC_performance_AUC}}
 \subfloat[Balanced Accuracy]{\includegraphics[width=0.45\linewidth]{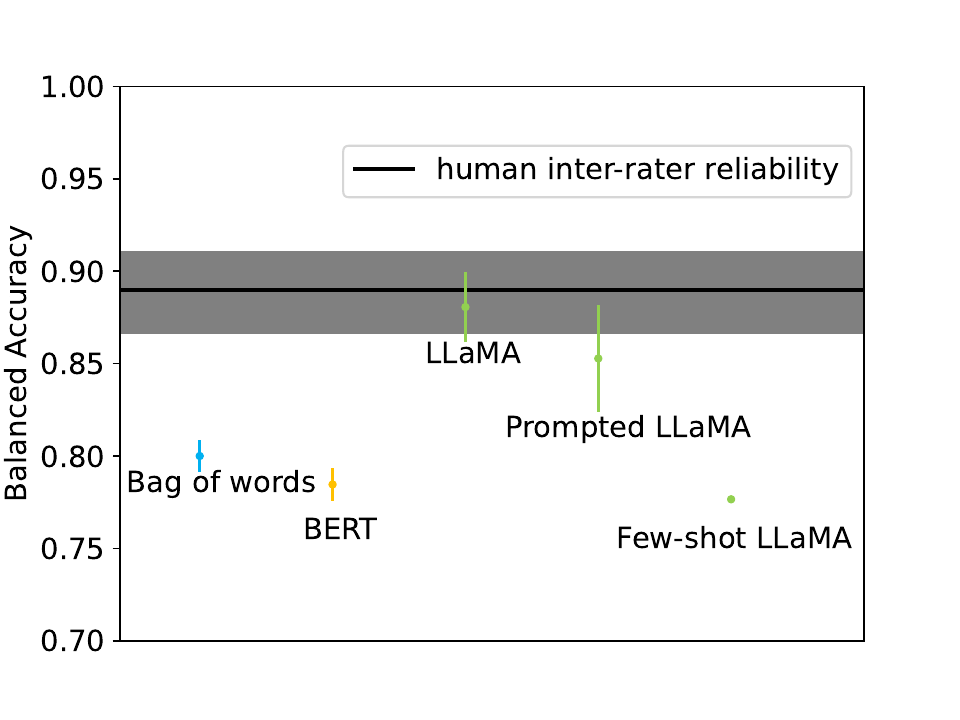}\label{QC_performance_bal_acc}}
 \hfill
\subfloat[Accuracy]{\includegraphics[width=0.45\linewidth]{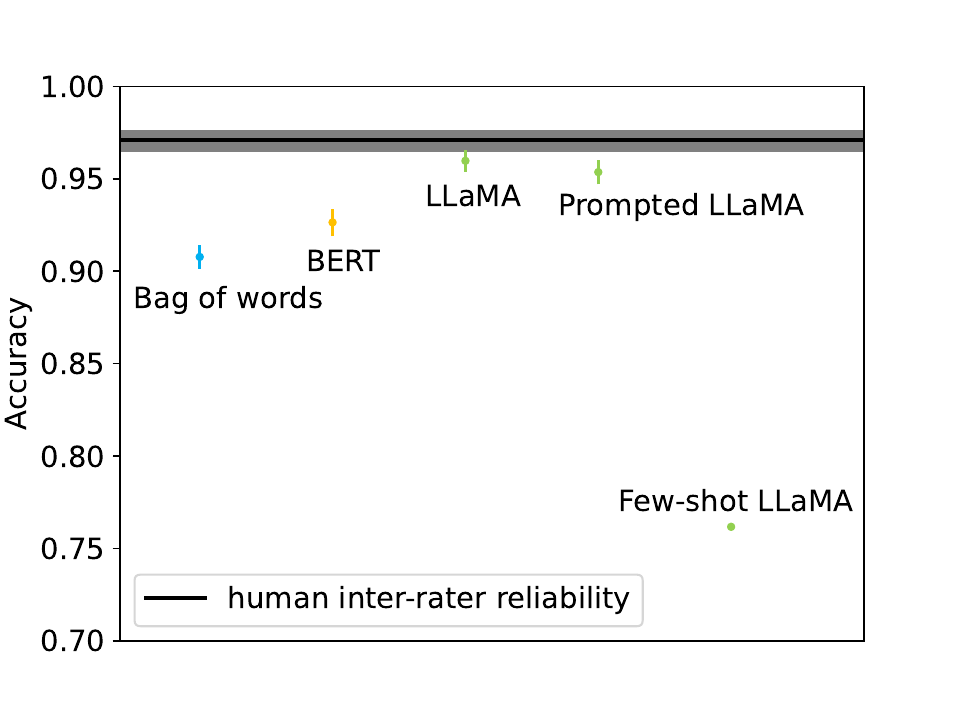}\label{QC_performance_acc}}
  \caption{Comparison of performance in applying the QC code. 
  Where possible, we indicate each metric calculated on the human inter-rater reliability data (black line with gray shading for representing one standard deviation across repeated samples of eight out of seventeen lab notes).~\label{QC_performance}}\vspace{-1em}
\end{figure*}
\begin{figure*}[t]
  \subfloat[AUC]{\includegraphics[width=0.45\linewidth]{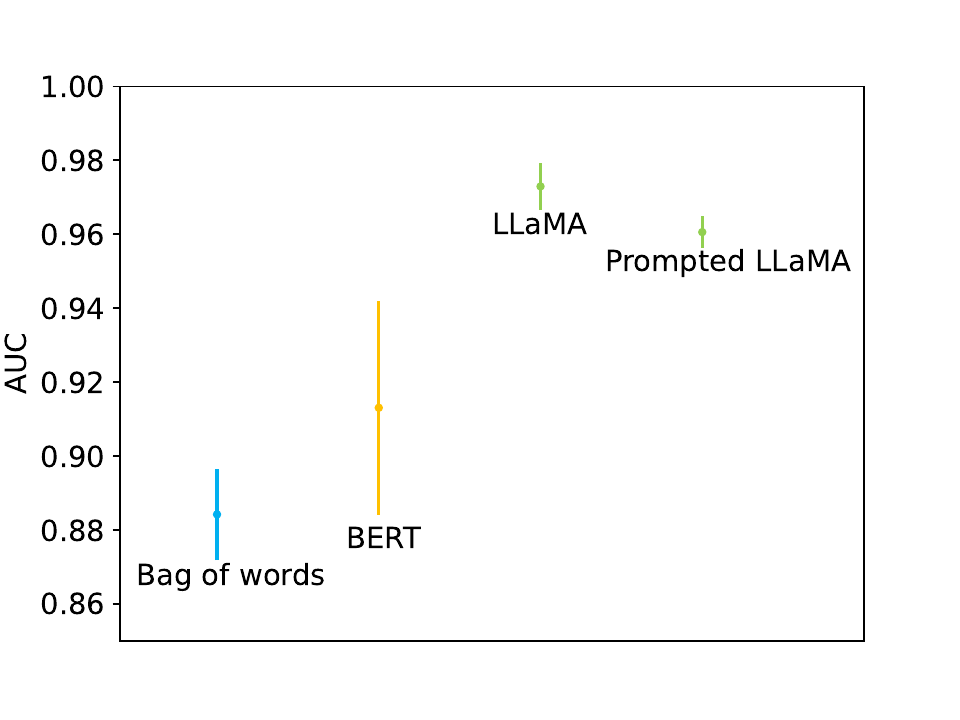}\label{PI_performance_AUC}}
 \subfloat[Balanced Accuracy]{\includegraphics[width=0.45\linewidth]{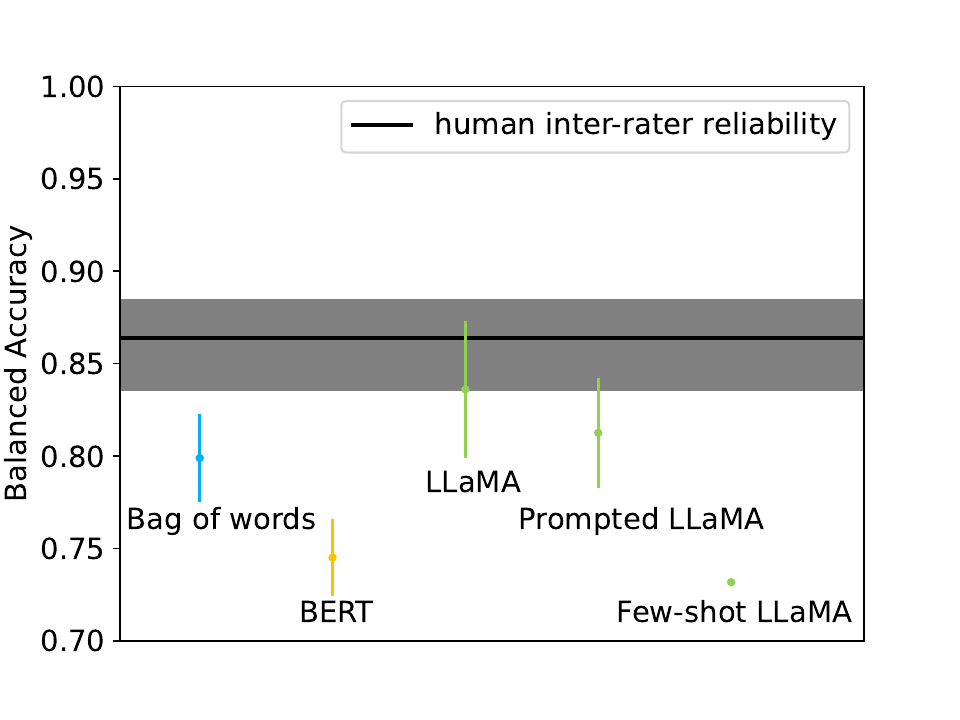}\label{PI_performance_bal_acc}}
 \hfill
 \subfloat[Accuracy]{\includegraphics[width=0.45\linewidth]{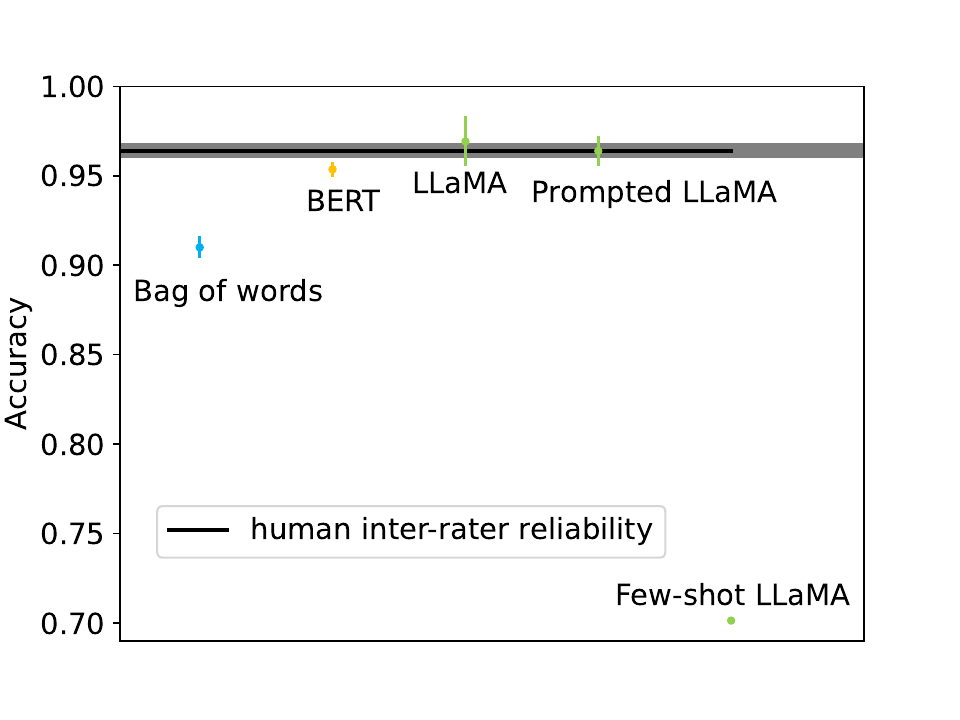}\label{PI_performance_acc}}
  \caption{Comparison of performance in applying the PI code. 
  Where possible, we indicate each metric calculated on the human inter-rater reliability data (black line with gray shading for uncertainty).~\label{PI_performance}}\vspace{-1em}
\end{figure*}

\begin{figure*}[t]
\subfloat[QC code]
{\includegraphics[width=0.5\linewidth]{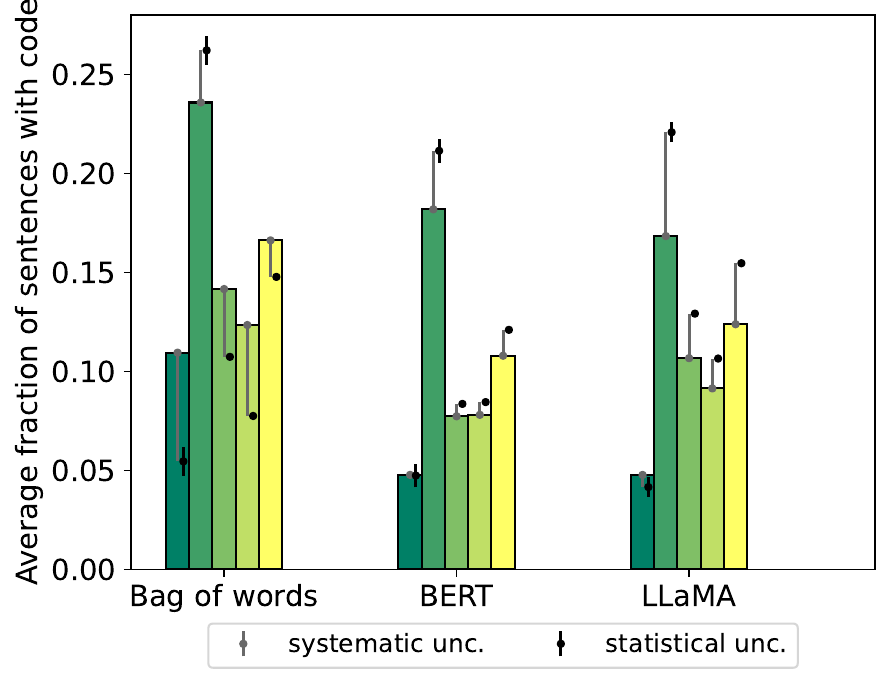}\label{barchart_QC}}
\subfloat[PI code]
{\includegraphics[width=0.5\linewidth]{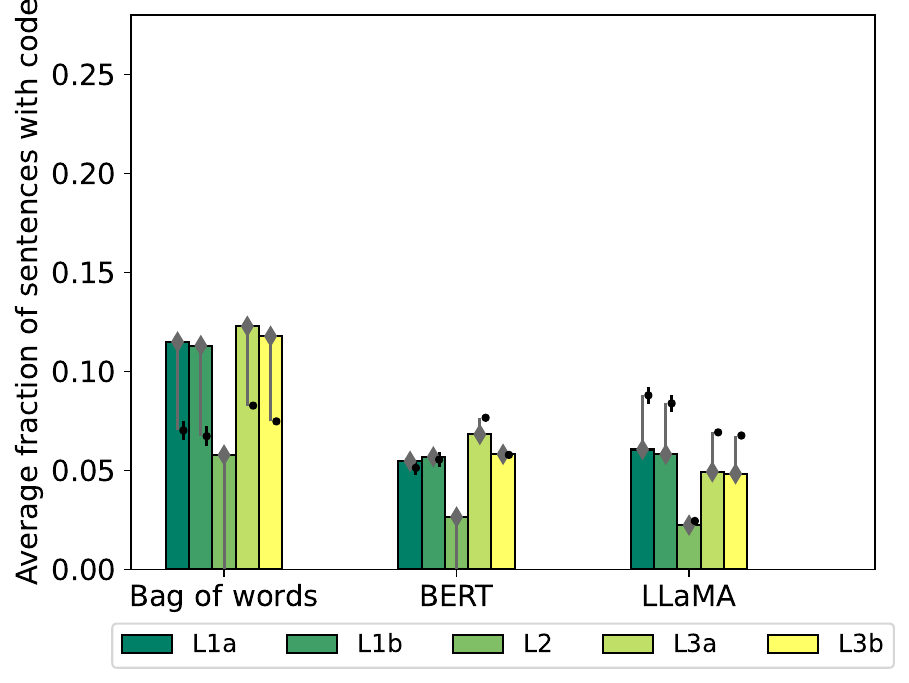}\label{barchart_PI}}
  \caption{Measurement outcomes of the average fraction of lab notes sentences across each class session, as measured by bag of words, BERT, and unprompted LLaMA. Statistical uncertainty (black error bars) is calculated as the standard deviation of the measurement across each lab notes document (labeled with lab unit number [e.g., L1] and session within that unit [a or b]) . Systematic uncertainty depicts the general direction and size of the correction needed to account for the measurement-dependent systematic uncertainty (gray error bars).~\label{barcharts}}\vspace{-1em}
\end{figure*}

\subsection{Performance}
In this section, we compare the average performance of the model types: bag of words, BERT, LLaMA with no prompt, prompted LLaMA, and \red{few-shot} LLaMA. Where possible, we compare the performance with human inter-rater reliability. The AUC metric is calculated using the output of the classifier model outputs, thus it is not possible to calculate AUC for the human inter-rater reliability. It is also not possible to calculate AUC for the \red{few-shot} LLaMA ``chatbot'' model, as this method also does not involve computing a model output. 

\subsubsection{Performance of models for the QC code} 


The two trained LLaMA models performed better than bag of words and BERT across all three metrics, making use of the greater computing power dedicated to training. The two trained LLaMA models have balanced accuracy that is almost comparable with the human inter-rater balanced accuracy (Fig.~\ref{QC_performance_bal_acc}), though the accuracy is less than the human inter-rater accuracy (Fig.~\ref{QC_performance_acc}). On the whole, these two LLaMA models were superior to other approaches. 

Surprisingly, the prompted LLaMA model performed similarly to the unprompted LLaMA model across all these metrics. This result suggests that the unprompted LLaMA model is near the ceiling of possible performance as it approaches the human inter-rater ceiling (we would not expect a model trained on human-coded data to outperform the inter-human metrics). Adding a prompt did not extend performance beyond this ceiling. 
In preliminary tests where the LLaMA model was not as well optimized, the prompted LLaMA model outperformed the unprompted model, indicating that the prompts can add value. 

Overall, the \red{few-shot} model significantly underperformed even the traditional bag of words approach.
The \red{few-shot} LLaMA model resulted in a balanced accuracy score comparable to that of BERT and bag of words and an accuracy score much lower than any of the trained models. We expect that the under-performance of the \red{few-shot} model arises at least in part from the fact that the generalized training of any \red{few-shot} model does not fully support the specificity of our coding scheme. This result, combined with the fact that the unprompted LLaMA model performed similarly to the prompted model, suggests that there is still a role for trained supervised models to be used in PER, rather than relying on prompt engineering alone. It is worth noting, however, that the model we used \red{for this few-shot analysis} is significantly smaller than those used in the most impressive commercial chat bots. Open-source LLMs of similar size are available, for example, a LLaMA model with the same architecture but 405 billion parameters (50 times larger than our current LLaMA model) is available. We do not have the resources to run it, but it may perform better than the \red{few-shot} model we used. 
 


Comparing the performance of bag of words to the performance of BERT illuminates an interesting trade-off. The balanced accuracy score for BERT is slightly worse than it is for bag of words (Fig.~\ref{QC_performance_bal_acc}), even though the accuracy score for BERT is slightly higher than it is for bag of words (Fig.~\ref{QC_performance_acc}). The higher bag of words balanced accuracy score does not mean that bag of words had better model output compared to BERT--because the AUC is higher for BERT--rather, it indicates that the bag of words prioritizes maximizing balanced accuracy more than BERT. This prioritization could happen by optimizing sensitivity, the rate of true positives over all positive predictions made by the model. Positive sentences are a minority case for the QC code, so admitting just a few more sentences as true positive sentences would result in large improvements to the sensitivity term in the balanced accuracy metric. The metric's specificity term, on the other hand, would be more robust to subtle changes in the algorithm because there are many more true negative sentences. This explanation is consistent with the lower accuracy score we see for bag of words. That is, prioritization of sensitivity may decrease the overall accuracy, as false positive sentences are admitted along with the true positives. 
When we examine the AUC metric, we see that BERT, the smallest LLM, represents a significant increase in performance over the traditional bag of words approach (Fig.~\ref{QC_performance_AUC}), which suggests that the BERT model's performed better than the bag of words model overall.


\subsubsection{Performance of models for the PI code}

Across the three metrics, differences across the models are not as apparent as they were for the QC code, though the unprompted LLaMA model again shows the best performance (Fig.~\ref{PI_performance_bal_acc}).
The prompted LLaMA model again performs similarly to the unprompted model on all three metrics, approaching the level of human inter-rater balanced accuracy and meeting the level of human inter-rater accuracy.
The \red{few-shot} LLaMA model has the lowest balanced accuracy (though nearly tied with BERT) and lowest accuracy (by a large margin). 
As with the QC code, the bag of words model appears better optimized for balanced accuracy compared to BERT while the BERT model appears better optimized for accuracy (Fig.~\ref{PI_performance_acc}).



\subsection{Research outcomes}
In this section, we describe the outcomes when three of our models (bag of words, BERT, and unprompted LLaMA) are used to answer an authentic research question: ``How does the prevalence of experimental skills change over the course of lab units in our introductory experimental physics course?'' For ease of interpretation, we have pared down the number of LLaMA models in this investigation to include just the model with the highest performance: the unprompted LLaMA model. We justify this decision because in a research study using LLaMA to analyze student data and report research outcomes, the outcomes would be based on the classifications from the best performing model with similar resources. 
We maintain the comparisons to bag of words and BERT because they require the least resources and so are the most accessible, despite their somewhat lower performance.


\subsubsection{Outcomes of the QC code}
In general, all three models (bag of words, BERT, and LLaMA) measure a similar trend for the prevalence of the QC code in students' lab notes over time (Fig.~\ref{barchart_QC}). For example, all models agree that the second lab session, L1b, has the highest prevalence (average fraction) of sentences with the QC code and L1a has the fewest. These data tell a story of the QC skill across the semester: in the first lab session, students include the fewest sentences demonstrating the QC skill; in the second session, there is scaffolding in the lab instructions that successfully prompts the students to engage the QC skill more frequently. After this session the scaffolding is removed and students do not engage the QC skill as often as they did in the second session, but there is an increase in use of this skill compared to the first session. Furthermore, the statistical uncertainty is smaller in the three later sessions, suggesting that there is more consistency across groups of students in the fraction of sentences they include demonstrating the QC skill. 

The specific measurements, however, differ across the three models, even though the models show similar trends. In each lab session, bag of words estimates higher average fraction values compared to the other two models before considering systematic effects. 
This result is consistent with our interpretation above that the bag of words model optimized for balanced accuracy, prioritizing sensitivity over specificity. Under this interpretation, the bag of words model's threshold would be set to assign the code to more of the sentences in the ambiguous region. This would have the double effect of correctly identifying a greater number of sentences that have the code (true positives) but also falsely labeling other sentences as containing the code (false positives). 

In general, the measurements from each model do not agree within uncertainties\red{, which are calculated using the methods described in Section~\ref{appendix}.} In some cases the systematics help to bring the three measurements more into alignment and in other cases these systematic corrections do not. For example, the systematics of the three L3b measurements point in the direction that brings the measurement closer to agreement with the other models' measurements.
The systematics for the three L1b measurements, in contrast, suggest that all three models could be underestimating. These discrepancies may arise from the fact that the systematics are calculated based on the rates of false positives and false negatives across all lab notes, when in reality the rates of false positives and false negatives likely vary across the different units. 
These results further demonstrate that reporting statistical and systematic uncertainty estimates associated with measurements made by models is relevant to assessing the value of a measurement for a claim in PER, but other factors, such as performance, must also be considered when making this assessment. Furthermore, the results motivate focusing research questions on comparisons and trends, rather than absolute value estimates.




\subsubsection{Outcomes of the PI code}

For the PI code, all three models again measure a similar trend, as with the QC code (Fig.~\ref{barchart_PI}). Specifically, the PI code applies to sentences across all lab sessions with a similar, low frequency (somewhere between 5 and 10\%), with the exception of L2, which has the lowest average fraction across all three models. This result again supports a story of engagement with the skill across the semester:  lab units that lasted two lab sessions provided opportunities for students to meaningfully propose iterations on their experimental procedures---the single session for L2 was comparatively less effective.

The specific measurements, however, differ across the three models, as we saw with the QC code. The direction of the systematic \red{uncertainty}, if not the magnitude, tends to point in the direction that would bring the measurement closer to an average of the three models' measurements \red{(see Section~\ref{appendix} for details on how uncertainty was calculated)}. For example, bag of words has higher estimates than the other models across all lab sessions. The systematic correction for all bag of words measurements is pointing downward and the systematic correction for all the LLaMA measurements is pointing upward, bringing the overall measurements closer together, though still not within statistical uncertainties of each other.

\section{Discussion}
When selecting a machine learning model for a study, there are multiple complex dimensions to consider. Most notably, there is a balance between performance and resources: in the benchmark tests commonly used to evaluate models in the machine learning literature, performance tends to improve with parameter count~\cite[e.g.,][]{devlin_bert_2019, touvron_llama_2023}. This principle generally held up in our results; BERT tended to perform better than bag of words and LLaMA tended to perform better than BERT.

Furthermore, the measurements made by the LLMs, both BERT and LLaMA, were more trustworthy in that these models classified more sentences correctly (that is, similarly to how humans would classify them). These results suggest it is worthwhile for physics education researchers to more regularly use LLMs in lieu of the more traditional machine learning approaches, such as bag of words. Choosing more cutting edge, resource-intensive LLMs, such as LLaMA, over simpler models, such as BERT, may be worthwhile in some cases, as our performance data show, but in other cases may not be worth the additional effort and resources. In our analysis, BERT did not approach human levels of inter-rater reliability for the QC code, but it approached inter-rater accuracy for the PI code (though not inter-rater balanced accuracy). Though the investment into LLaMA was worthwhile for our task of identifying skills in lab notes (because it resulted in improved performance compared to the lower-resource models), we expect that the additional investment into LLaMA may not be worthwhile for other tasks where BERT's performance approaches human levels of inter-rater reliability (e.g. a task involving a simpler coding scheme). In addition, all three types of models were able to identify similar trends when answering a sample research question. For certain research purposes, this may be sufficient, questioning the need for higher resources.

Large language models, and neural networks broadly, are also highly complex systems (in a non-linear dynamics sense), with many factors that may impact performance in both predictable and unpredictable ways. 
As an example of this unpredictability, the models we present here are the results of several iterations of hyper-parameter adjustments.  
In earlier iterations, the performance of the LLaMA models we tested was lower than the performance of BERT. 
The broader machine learning community does not know the optimal set of hyper-parameters for every problem, though BERT has been studied for longer than LLaMA.
This, combined with the much shorter runtime of training BERT made it significantly easier to choose hyper-parameters.   
With LLaMA, results may be unpredictable especially if one does not have the resources to run hyper-parameter optimization algorithms that involve training models many times.  
In our case, allowing LORA to insert more low rank layers~\cite{mamba_llama3_nodate} significantly improved the accuracy of LLaMA, without this we would not be able to present a LLaMA model that outperformed BERT.  

Importantly, we found that models fine-tuned with training data significantly outperformed the \red{few-shot} model. There is a trend in PER towards relying on prompt engineering alone, with no task-specific fine-tuning, when using LLMs to analyze data~\cite[e.g.,][]{kortemeyer_toward_2023,kortemeyer_could_2023, polverini_performance_2024}. Our data show that for theory-driven classification tasks in PER, supervised trained models may be more useful than \red{few-shot}. This result holds at all levels we investigated: traditional bag of words, BERT, and LLaMA. Supervised trained models for theory-driven classification tasks may involve hand-coding a large dataset, as we do here, though other approaches, such as training the classifier using only representative sentences in a text embedding, may be just as promising~\cite{odden_using_2024}.

It is interesting that BERT, which was trained with the idea of being modified for classification, did not perform as well as LLaMA, which was trained for text generation. Our results suggest that generative models can also perform well for non-generative research tasks, such as automated application of a coding scheme. 

In examining outcomes to an education research question, we saw that the measurements of prevalence of the QC and PI codes varied by model. The overall trends, however, were consistent between the models. The results support interpreting machine learning outputs by incorporating uncertainties (both statistical and systematic) and focusing on trends, not specific values. 

\red{In this paper we restricted ourselves to open-source models that we could run locally because future studies in PER may need to keep many forms of human subjects data private and secure. This restriction need not apply to all PER studies, however. Some academic research institutes, for example, have contracts with providers of online LLM resources that include provisions that their research data will not be used for training. Additionally, some studies that use online LLM resources may be ethically permissible if all data are de-identified.}

Our comparison of the traditional bag of words approach to BERT and the much larger LLaMA has provided interesting insights into recommended natural language processing methods in PER. However, large language model innovations are changing quickly. \red{The largest model we used was the 8-billion parameter version of LLaMA 3, which is much smaller than the latest proprietary models, such as GPT-4. Open-source LLMs of comparable size and performance to GPT-4 are available but we do not have the resources to run them locally. Outsourcing the processing is possible through cloud computing but would have contravened our purpose to run all analysis locally and securely.} The effectiveness \red{and ethics} of different LLMs for PER studies will likely be a moving target. The analyses here, therefore, can serve as a template for decision making around future use studies.

\begin{acknowledgments}
MF and AD were supported by the SciAI Center, funded by the Office of Naval Research under Grant Number N00014-23-1-2729. This work was also supported by the Imperial-Cornell Education Seed Fund. We are grateful to Jasmine Ajaz, Jonte Catton, Le-Qi Tang, and Sabrina McDowell, who helped with coding all the lab notes.
\end{acknowledgments}

\appendix

\section{Appendix}\label{appendix}
In the main text, we report systematic and statistical uncertainty alongside our measurements of the average fraction of sentences per lab notes document with one of the codes, PI or QC. Here, we describe how the statistical and systematic uncertainty were calculated. 

\subsection{Statistical Uncertainty}
To make the measurements in fig.~\ref{barchart_QC}, we use the fraction of sentences in \textit{each} lab notes document than contain the code, ${f_1, f_2, ... f_i ... f_N}$, where $N$ is the number of lab notes documents in the given unit and session of the lab course. The height of each bar, the mean fraction for a given unit and session of the lab course, is the mean of this set of $f_i$'s. The statistical uncertainty is the standard uncertainty of the mean of this set of fractions. 

\subsection{Systematic Uncertainty}
\begin{figure*}[t]
  \subfloat[Systematic effects for bag of words]{\includegraphics[width=0.3\linewidth]{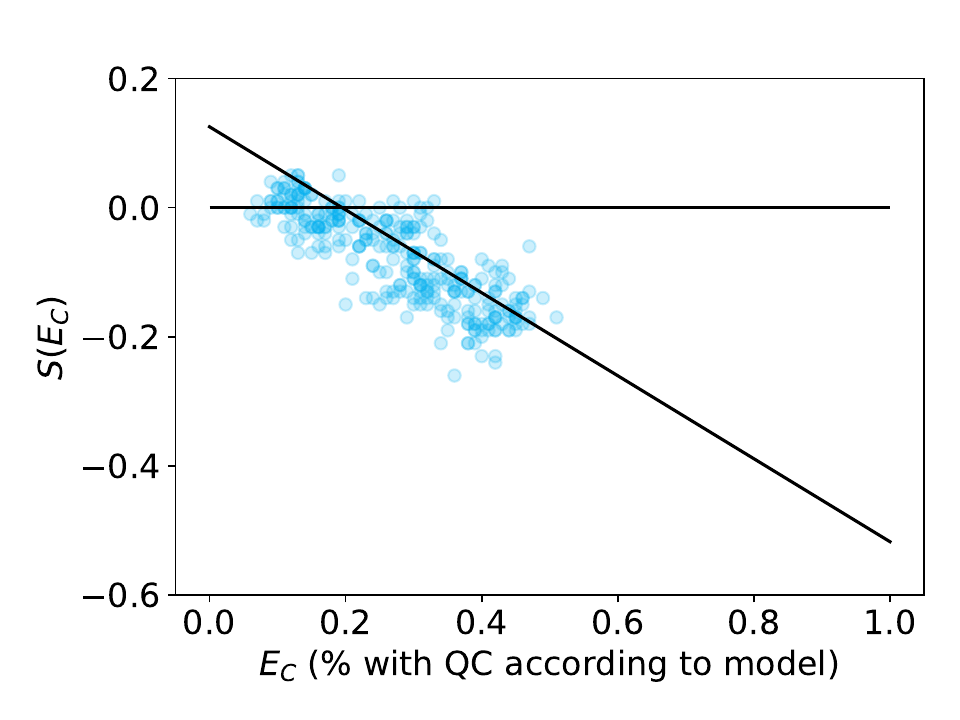}}
 \subfloat[Systematic effects for BERT model]{\includegraphics[width=0.3\linewidth]{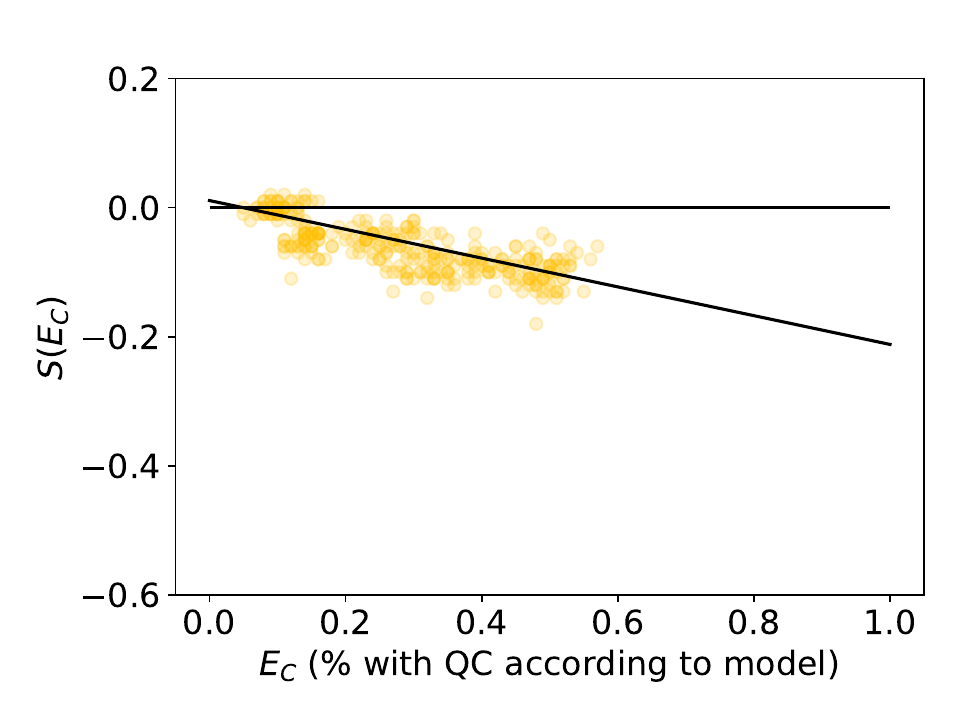}}
 \subfloat[Systematic effects for unprompted LLaMA model]{\includegraphics[width=0.3\linewidth]{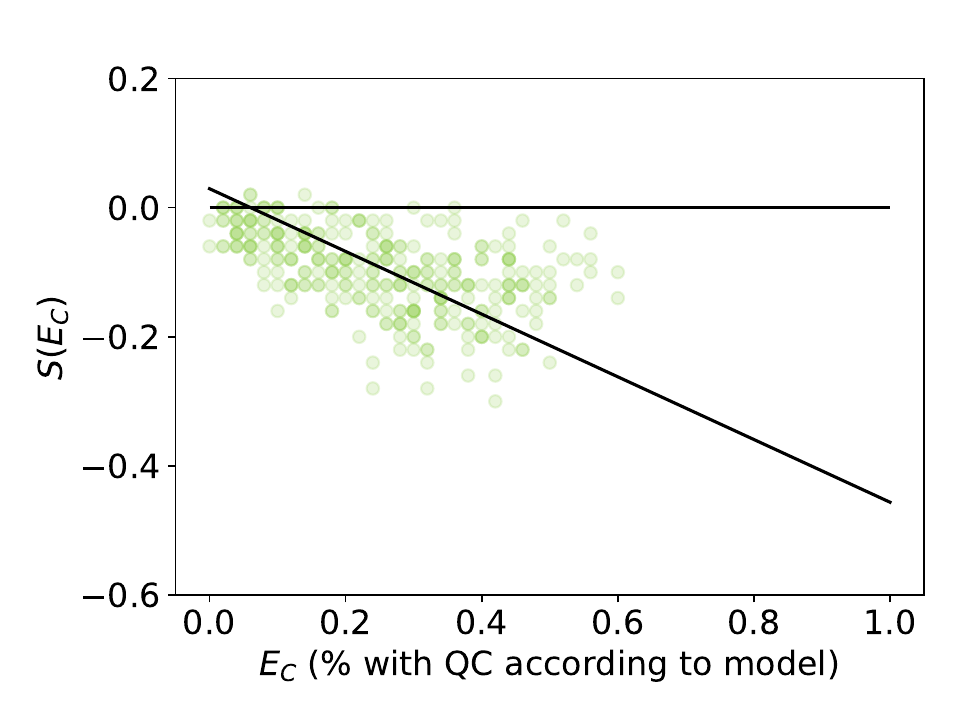}}
  \caption{Computing the systematic uncertainty in Fig.~\ref{barchart_QC} (QC code outcomes): given a value of $E_C$, we compute the systematic uncertainty using Eqn.~\ref{eqn:SEC}. Scatterplot points around the line of best fit represent random sampling for samples at different code frequencies. ~\label{systematics_QC}}\vspace{-1em}
\end{figure*}

\begin{figure*}[t]
  \subfloat[Systematic effects for bag of words]{\includegraphics[width=0.3\linewidth]{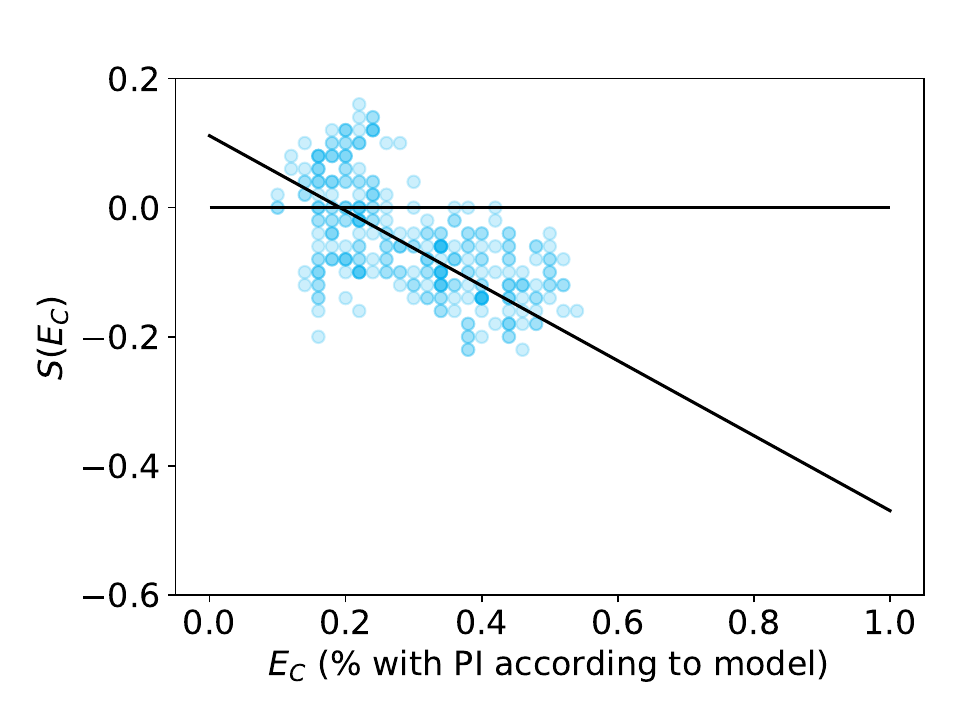}}
 \subfloat[Systematic effects for BERT model]{\includegraphics[width=0.3\linewidth]{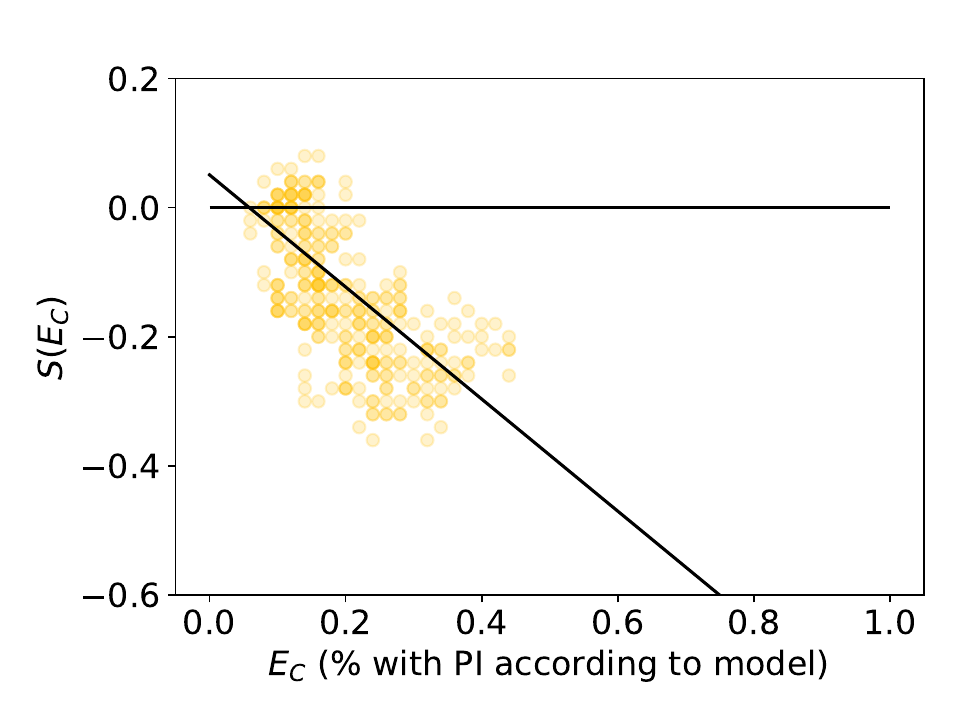}}
 \subfloat[Systematic effects for unprompted LLaMA model]{\includegraphics[width=0.3\linewidth]{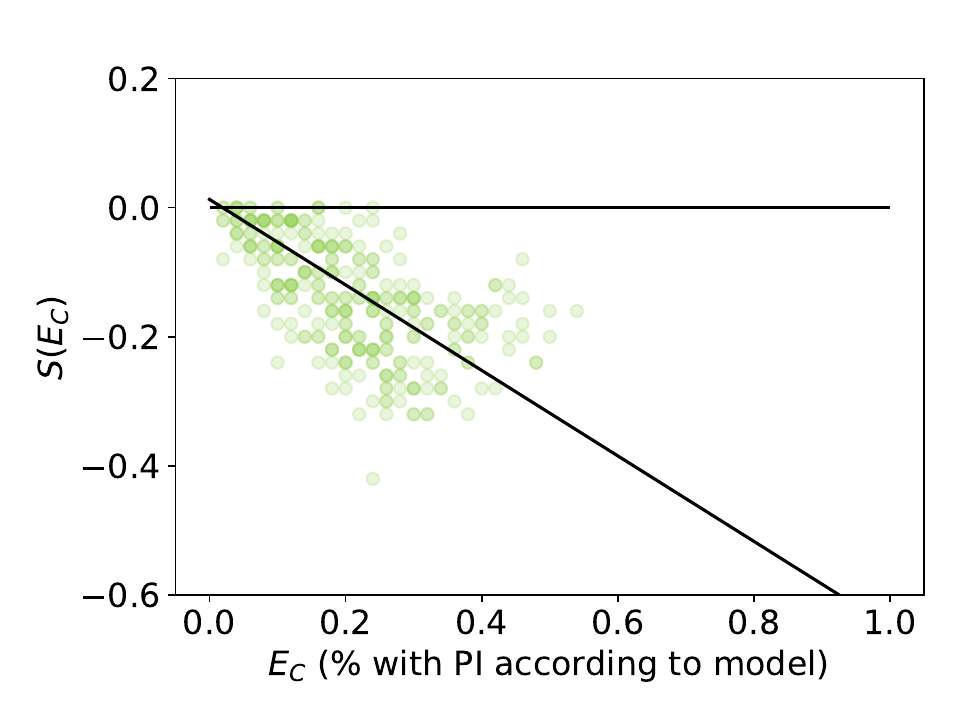}}
  \caption{Computing the systematic uncertainty in Fig.~\ref{barchart_PI} (PI code outcomes): given a value of $E_C$, we compute the systematic uncertainty using Eqn.~\ref{eqn:SEC}. Scatterplot points around the line of best fit represent random sampling for samples at different code frequencies.~\label{systematics_PI}}\vspace{-1em}
\end{figure*}

We compute the systematic uncertainty through two methods, a closed form solution and through random sampling. 

\subsubsection{Closed form solution}
We use the confusion matrix for a model's predictions on the hold out test set (N = 1242 sentences) to compute the closed form solution.
Given a labeled test set of size N, we measure the model output on the data and calculate a confusion matrix given by Table \ref{confusion}, 
where $TP$, $FP$, $FN$ and $TN$ stand for True Positive, False Positive, False Negative and True Negative respectively. We consider the model to be accurate when it agrees with a human coder.

\begin{table}
\caption{The confusion matrix}\label{confusion}
\begin{tabular}{l|l|c|c|}
\multicolumn{2}{c}{}&\multicolumn{2}{c}{Human coder}\\
\cline{3-4}
\multicolumn{2}{c|}{}&Positive&Negative\\
\cline{2-4}
\multirow{2}{*}{Model}& Positive & $TP$ & $FP$\\
\cline{2-4}
& Negative & $FN$ & $TN$\\
\cline{2-4}
\end{tabular}
\end{table}

Next, we apply the model on a new set of sentences. We define $E_H$ as the percentage of sentences with the code according to the human coder's estimate, which is unknown for unlabeled data. Given this, we expect our machine classifier to measure the percentage of the sentences containing the code,$E_c$, as 

\begin{equation}
    E_c = E_H P_1 +(1-E_H)P_2.
    \label{ec}
\end{equation}

In words, this is:  The measured percentage of sentences containing the code will be equal to the probability that a randomly selected sentence contains the code times the probability that a sentence that contains the code is classified correctly by the classifier ($P_1$), plus the probability that a randomly selected sentence does not contain the code times the probability that a sentence that does not contain the code is incorrectly classified by the classifier ($P_2$).

We can calculate these two probabilities from the confusion matrix as  

\begin{equation}
    P_1= \frac{TP}{TP+FN},
\end{equation}

\begin{equation}
    P_2= \frac{FP}{FP+TN}.
\end{equation}

We make the assumption that $P_1$ and $P_2$ do not change at different levels of code prevalence in the dataset (that is, different values of $E_H$). In other words, neither the probability that a sentence is misclassified given that it does contain the code, nor the probability that a sentence is misclassified given that it does not contain the code, change within a single model and data source. 

Using Eqn.~\ref{ec} to solve for $E_H$, we find that
\begin{equation}
    E_H=\frac{E_C-P_2}{P_1-P_2}.
\end{equation}
This is equivalent to Eqn. 5 in ~\cite{ott_estimating_2012}.

If we want the expected systematic uncertainty as a function of $E_C$, 
\begin{equation}
    S(E_C)=E_C-E_H=E_C-\frac{E_C-P_2}{P_1-P_2}
\end{equation}
or 
\begin{equation}
    S(E_C)=E_C\left(1-\frac{1}{P_1-P_2}\right)+\frac{P_2}{P_1-P_2}.\label{eqn:SEC}
\end{equation}

\subsubsection{Random sampling}
The method to compute systematic uncertainty through random sampling involves taking samples of validation data and fixing the code frequency for different samples. We compute the systematic difference between the model's prediction of code frequency in the sample and the human coders' prediction of code frequency in the sample. We tend to observe that this systematic difference relates to the frequency of the code with a negative-sloped linear relationship. We fit a line to the relationship between the systematic difference and the frequency of the code in the validation data, and we infer that a similar line applies to uncoded data. For more methodological details, we refer the reader to~\cite{fussell_method_2024}.

\subsubsection{Bringing systematic uncertainty methods together}
We plot the relationship between the percent of sentences with the code, $E_C$ (also written as \%QC and \%PI for the QC and PI codes respectively), and $S(E_C)$ for each model and code in Figs.~\ref{systematics_QC} and ~\ref{systematics_PI}. The lines represent the solution to Eqn.~\ref{eqn:SEC} with the values of $P_1$ and $P_2$ specific to the confusion matrix for that model and code. The points in the scatterplot come from the method to compute systematic uncertainty through random sampling, blue for the bag of words model, yellow for the BERT model, and green for the unprompted LLaMA model.

\bibliography{Zotero_Library}

\end{document}